\documentclass[12pt,prd,showpacs,tightenlines,nofootinbib]{revtex4}
\usepackage{bm}
\usepackage{graphics}
\usepackage{rotating}
\usepackage{epsfig}
\begin{document}
\title{\begin{flushright}{\rm\normalsize HU-EP-04/24}\end{flushright}
Semileptonic decays of doubly heavy baryons in the
relativistic quark model}
\author{D. Ebert}
\affiliation{Institut f\"ur Physik, Humboldt--Universit\"at zu Berlin,
Newtonstr. 15, D-12489  Berlin, Germany}
\author{R. N. Faustov}
\author{V. O. Galkin}
\affiliation{Institut f\"ur Physik, Humboldt--Universit\"at zu Berlin,
Newtonstr. 15, D-12489 Berlin, Germany}
\affiliation{Russian Academy of Sciences, Scientific Council for
Cybernetics, Vavilov Street 40, Moscow 117333, Russia}
\author{A. P. Martynenko}
\affiliation{Institut f\"ur Physik, Humboldt--Universit\"at zu Berlin,
Newtonstr. 15, D-12489 Berlin, Germany}
\affiliation{Samara State University, Pavlov Street 1, Samara 443011,
  Russia} 

\begin{abstract}
Semileptonic decays of doubly heavy baryons are studied in the framework of
the relativistic quark model. The doubly heavy baryons are treated in
the quark-diquark approximation. The transition amplitudes of heavy diquarks
$bb$ and $bc$ going respectively to $bc$ and $cc$  are explicitly expressed
through the overlap integrals of the diquark wave functions in the
whole accessible kinematic range. The relativistic 
baryon wave functions of the quark-diquark bound system  are used for
the calculation of the transition matrix elements, the Isgur-Wise function
and decay rates  in the heavy quark limit. 
\end{abstract}


\maketitle

\section{Introduction}

The description of doubly heavy baryon properties acquires in the last years
the status of actual physical problem which can be studied
experimentally. The appearance of experimental data on  $B_c$ mesons
\cite{GKL}, heavy-light baryons \cite{UFN} stimulates the
investigation of heavy quark bound states and can help in discriminating
numerous quark models. 
Recently first experimental indications of the existence of doubly charmed
baryons were published by SELEX \cite{selex}. Although these data need
further 
experimental confirmation and clarification it manifests that in the
near future the mass spectra and decay rates of doubly heavy baryons
will be measured. This gives additional grounds
for the theoretical investigation of the doubly heavy baryon properties.
The success of
the heavy quark effective theory (HQET) \cite{MN} in predicting 
properties of the heavy-light $q\bar Q$ mesons ($B$ and $D$)
suggests to apply these methods to heavy-light baryons, too.
The semileptonic decays of heavy hadrons present also an important tool
for determining the parameters of the Cabibbo-Kobayashi-Maskawa
(CKM) matrix. 

The relativistic quark model has been remarkably
successful in describing the observed hadronic states and their decay rates
\cite{p1,r1,r2,r3}. This includes the heavy-light mesons and heavy
quarkonia. The quark model
predicts the existence of doubly heavy baryons containing two heavy
quarks ($cc$, $bc$, $bb$) and one light quark ($u,d,s$). The energies
necessary 
to produce these particles are already reached. The main difficulty remains
in their reconstruction since these states have in general a large number
of decay modes and thus high statistics is required \cite{GKL}.

Doubly heavy baryons occupy a special position among existing
baryons because they can be studied in the quark-diquark approximation
and the two-particle bound state methods can be applied. 
The two heavy quarks
compose in this case a bound diquark system in the
antitriplet colour state which serves as a localized colour
source. The light quark $q$ is orbiting around this heavy source at a
distance much larger ($\sim1/m_q$) than the source size
($\sim2/m_Q$). The estimates of the light quark velocity in these
baryons show that its value is $v/c\sim
0.7\div 0.8$ and the light quark should be treated fully
relativistically.
Thus the doubly heavy baryons look effectively like a
two-body bound system and strongly resemble the heavy-light $B$ and
$D$ mesons \cite{bali,isgur}. Then the HQET expansion in the inverse heavy
diquark mass can be performed.  We used a similar approach for
the calculation of the mass spectra of 
doubly heavy baryons \cite{r3}. The ground state baryons with two heavy
quarks can be composed from a compact doubly heavy diquark of spin 0
or 1 and a light quark. According to the Pauli principle the diquarks
$(bb)$ or $(cc)$ have the spin 1 whereas diquark $(bc)$ can have both
the spin 0 and 1. 

There exists already a number of papers devoted to studying both
the mass spectra of doubly heavy baryons and their decay rates
\cite{UFN,p1,KO,SSG,Ivanov,IN,Lozano,Guo,AO,WS,GW,CC,AF,Hussain}.
Whereas the results for the mass spectra are in agreement, the
calculations of exclusive semileptonic decays lead to essentially
different values for the decay rates \cite{Ivanov,Lozano,Guo,AO}
obtained with the help of the Bethe-Salpeter equation, QCD sum rules and
relativistic three quark model. 

\begin{figure}[htbp]
  \centering
  \includegraphics{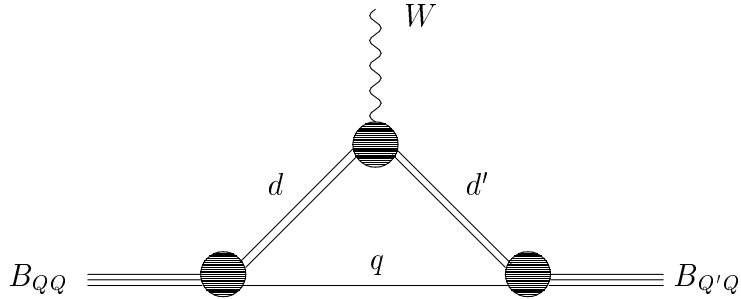}
  \caption{Weak transition matrix element of the doubly heavy baryon in the
  quark-diquark approximation.}
  \label{mdiag}
\end{figure}

Here we study semileptonic decay rates of doubly heavy baryons using the
relativistic quark model in the quark - diquark approximation. The
covariant expressions for the semileptonic decay amplitudes
of the baryons with the spin 1/2, 3/2 are obtained in the limit $m_c,
m_b\to\infty$ and compared  with the predictions of HQET. 
The calculation of semileptonic decays of doubly heavy baryons $(bbq)$
or $(bcq)$ to doubly heavy baryons $(bcq)$ or $(ccq)$ 
can be divided into two steps (see Fig.~\ref{mdiag}). The first step
is the study of form factors of the weak transition between initial
and final doubly heavy diquarks. The second one consists in the
inclusion of the light quark in order to compose a baryon with
spin 1/2 or 3/2. 

The paper is organized as follows. In Sec.~\ref{rqm} we describe our
relativistic quark model and present predictions for the masses of
ground state heavy diquarks and doubly heavy baryons.  We
apply our model to the investigation of the heavy diquark
transition matrix elements in Sec.~\ref{hdf}. The transition
amplitudes of heavy diquarks are explicitly expressed in a covariant
form through the overlap integrals of the diquark wave functions. The
obtained general expressions reproduce in the appropriate limit  the
predictions of heavy quark symmetry. Section~\ref{dhbt} is devoted to
the construction of  transition matrix elements between doubly heavy baryons
in the quark-diquark approximation. The corresponding Isgur-Wise
function is determined. In Sec.~\ref{sdr} semileptonic decay rates of
doubly heavy baryons are calculated in the nonrelativistic limit for
heavy quarks. Our conclusions are given in Sec.~\ref{conc}.

\section{Relativistic quark model}  
\label{rqm}

In the quasipotential approach and quark-diquark picture of doubly
heavy baryons the interaction of two heavy quarks in a diquark and of
the light 
quark with a heavy diquark in a baryon are described by the
diquark wave function ($\Psi_{d}$) of the bound quark-quark state
and by the baryon wave function ($\Psi_{B}$) of the bound quark-diquark
state, respectively.  These wave functions satisfy the two-particle
quasipotential equation \cite{lt,F} of the Schr\"odinger type \cite{4}
\begin{equation}
\label{quas}
{\left(\frac{b^2(M)}{2\mu_{R}}-\frac{{\bf
p}^2}{2\mu_{R}}\right)\Psi_{d,B}({\bf p})} =\int\frac{d^3 q}{(2\pi)^3}
 V({\bf p,q};M)\Psi_{d,B}({\bf q}),
\end{equation}
where the relativistic reduced mass is
\begin{equation}
\mu_{R}=\frac{E_1E_2}{E_1+E_2}=\frac{M^4-(m^2_1-m^2_2)^2}{4M^3},
\end{equation}
and the center of mass energies of particles on the
mass shell $E_1$, $E_2$ are given by
\begin{equation}
\label{ee}
E_1=\frac{M^2-m_2^2+m_1^2}{2M}, \quad E_2=\frac{M^2-m_1^2+m_2^2}{2M}.
\end{equation}
Here $M=E_1+E_2$ is the bound state mass (diquark or baryon),
$m_{1,2}$ are the masses of heavy quarks ($Q_1$ and $Q_2$) which form
the diquark or of the heavy diquark ($d$) and light quark ($q$) which form
the doubly heavy baryon ($B$), and ${\bf p}$  is their relative momentum.  
In the center of mass system the relative momentum squared on mass shell 
reads
\begin{equation}
{b^2(M) }
=\frac{[M^2-(m_1+m_2)^2][M^2-(m_1-m_2)^2]}{4M^2}.
\end{equation}

The kernel 
$V({\bf p,q};M)$ in Eq.~(\ref{quas}) is the quasipotential operator of
the quark-quark or quark-diquark interaction. It is constructed with
the help of the
off-mass-shell scattering amplitude, projected onto the positive
energy states. Here we closely follow the
similar construction of the quark-antiquark interaction in heavy mesons
which were extensively studied in our relativistic quark model
\cite{egf,pot}. For
the quark-quark interaction in a diquark we use the relation
$V_{QQ}=V_{Q\bar Q}/2$ arising under the assumption about the octet
structure of the interaction  from the difference of the projection
onto $QQ$ and $Q\bar Q$  colour states. 
The quasipotential of the quark-antiquark interaction 
is the sum of the usual one-gluon exchange term and the confining part
which is the mixture
of long-range vector and scalar linear potentials, where
the vector confining potential contains the Pauli terms.  
The explicit expressions for these quasipotentials are given in
Ref.~\cite{r3}.  
The quark masses have the following values
$m_b=4.88$~GeV, $m_c=1.55$ GeV, $m_s=0.50$ GeV, $m_{u,d}=0.33$ GeV.  

We calculated in the framework of the relativistic quark model the mass
spectra of heavy diquarks and doubly heavy baryon masses in the
quark-diquark approximation in Ref.~\cite{r3}. The masses of the
ground state  axial vector  diquarks were found to be
$M_{cc}^{AV}=3.226$~GeV, $M_{bb}^{AV}=9.778$~GeV,
$M_{bc}^{AV}=6.526$~GeV, and the mass of the scalar diquark
$M_{bc}^S=6.519$~GeV. The calculated masses of the ground state
doubly heavy baryons are listed in Table~\ref{tab:gmass}. 

\begin{table}
\caption{\label{tab:gmass} Mass spectrum of ground states of doubly
  heavy baryons (in GeV) \cite{r3}. $\{QQ\}$ denotes the diquark in
  the axial 
  vector state and $[QQ]$ denotes the diquark in the scalar state.}
\begin{ruledtabular}
\begin{tabular}{cccc}
Baryon&
Quark content&$J^P$&Mass\\  
\hline
$\Xi_{cc}$  &$\{cc\}q$&$1/2^+$&3.620\\
$\Xi_{cc}^*$&$\{cc\}q$&$3/2^+$&3.727\\
$\Omega_{cc}$&$\{cc\}s$&$1/2^+$&3.778\\
$\Omega_{cc}^*$&$\{cc\}s$&$3/2^+$&3.872\\
$\Xi_{bb}$  &$\{bb\}q$&$1/2^+$&10.202\\
$\Xi_{bb}^*$ &$\{bb\}q$&$3/2^+$&10.237\\
$\Omega_{bb}$&$\{bb\}s$&$1/2^+$&10.359\\
$\Omega_{bb}^*$&$\{bb\}s$&$3/2^+$&10.389\\
$\Xi_{cb}$  &$\{cb\}q$&$1/2^+$&6.933\\
$\Xi'_{cb}$  &$[cb]q$&$1/2^+$&6.963\\
$\Xi_{cb}^*$ &$\{cb\}q$&$3/2^+$&6.980\\
$\Omega_{cb}$  &$\{cb\}s$&$1/2^+$&7.088\\
$\Omega'_{cb}$  &$[cb]s$&$1/2^+$&7.116\\
$\Omega_{cb}^*$ &$\{cb\}s$&$3/2^+$&7.130\\
\end{tabular}
\end{ruledtabular}
\end{table}

\section{Heavy diquark transition form factors}
\label{hdf}

The form factors of the subprocess $d(QQ_s)\to d'(Q'Q_s)e\bar\nu$ where one
heavy quark $Q_s$ is a spectator are determined by the weak decay of
the active heavy quark $Q\to Q'e\bar\nu$. The local effective
Hamiltonian is given by 
\begin{equation}\label{heff}
H_{eff}=\frac{G_F}{\sqrt{2}}V_{QQ'}\left(\bar
  Q'\gamma_\mu(1-\gamma_5)Q\right)\left(\bar
  l\gamma_\mu(1-\gamma_5)\nu_e\right), 
\end{equation}
where $G_F$ is the Fermi constant and $V_{QQ'}$ is the CKM matrix element.
In the relativistic quark model 
the transition matrix element between two diquark states is determined
by the contraction of the wave functions $\Psi_d$ of the initial and 
final diquarks with the two particle vertex function $\Gamma$ \cite{F}
\begin{equation}\label{tmm}
\langle d'(Q)|J_\mu^W|d(P)\rangle=\int\frac{d^3{ p}\, d^3{
    q}}{(2\pi)^6}\bar\Psi^{\lambda\sigma}_{d',Q}({\bf q}) 
\Gamma_\mu^{\lambda\sigma,\rho\omega}({\bf p},{\bf
    q})\Psi_{d,P}^{\rho\omega}({\bf p}), \qquad
    \lambda,\sigma,\rho,\omega=\pm\frac12.   
\end{equation}
Here we denote mass, energy and velocity  of the initial
 diquark
($Q_b Q_s$, index $b$ stands for the initial active  
quark and index $s$ for the spectator) by $M_i$, $E_i=M_i v^0$ and
$v=P/M_i$, and mass, energy and 
velocity of the final diquark  ($Q_a Q_s$, index $a$ means the final
 active quark) by
$M_f$, $E_f=M_f v'^{0}$ and $v'=Q/M_f$.

\begin{figure}
  \centering
  \includegraphics{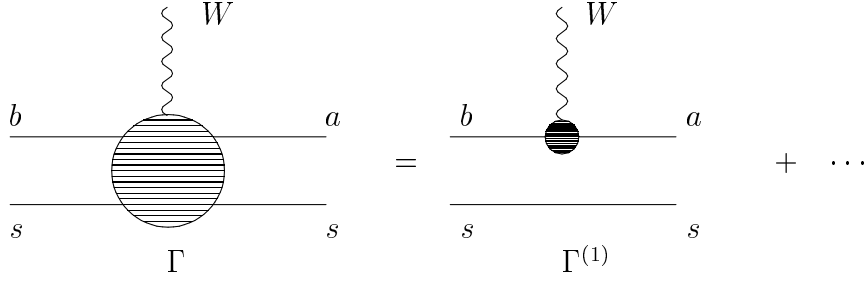}
  \caption{The leading order contribution $\Gamma^{(1)}$ to the
  diquark vertex function $\Gamma$.}
  \label{diag}
\end{figure}

The leading contribution to the vertex function $\Gamma_\mu$ comes
from the diagram in  Fig.~\ref{diag} \cite{F,FG,FGM}
(we explicitly show spin indices)
\begin{equation}\label{gamma1}
\Gamma_\mu^{\lambda\sigma,\rho\omega}({\bf p},{\bf q})
=\Gamma_\mu^{(1)}=\bar u_a^\lambda({\bf q}_1)\gamma_\mu
(1-\gamma_5)u_b^\rho({\bf p}_1)\bar u_s^\sigma({\bf
  q}_2)u_s^\omega({\bf p}_2) 
(2\pi)^3\delta({\bf p}_2-{\bf q}_2)\delta^{\sigma\omega},
\end{equation}
where the Dirac spinors are
\begin{equation}
\label{spinor}
u^\lambda({p})=\sqrt{\frac{\epsilon(p)+m}{2\epsilon(p)}}
\left(\begin{array}{c}
1\\ \displaystyle\frac{\mathstrut(\bm{\sigma}{\bf p})}
{\mathstrut\epsilon(p)+m}
\end{array}\right)
\chi^\lambda, \qquad \epsilon(p)=\sqrt{{\bf p}^2+m^2}, \quad
\lambda=\pm\frac12. 
\end{equation}
 Relativistic four-momenta of the particles 
in the initial and final states are defined as follows
\begin{eqnarray}\label{pqm}
p_{1,2}&=&\epsilon_{1,2}(p)v\pm\sum_{i=1}^3n^{(i)}(v)p^i,
\quad v=\frac{P}{{\cal M}_i},\quad {\cal
  M}_i=\epsilon_1(p)+\epsilon_2(p),\cr 
q_{1,2}&=&\epsilon_{1,2}(q)v'\pm\sum_{i=1}^3n^{(i)}(v')q^i,
\quad v'=\frac{Q}{{\cal M}_f},\quad {\cal M}_f=\epsilon_1(q)+\epsilon_2(q)
\end{eqnarray}
and $n^{(i)}$ are three four vectors defined by
\begin{displaymath}
n^{(i)}(v)=\left\{v^i, \delta^{ij}+\frac{1}{1+v^0}v^iv^j\right\}.
\end{displaymath}
After making necessary transformations,
the expression for $\Gamma$ should be continued in
${\cal M}_i$  and ${\cal M}_f$ to
the values of initial $M_i$ and final $M_f$ bound state masses.

The transformation of the bound state wave functions from the rest frame
to the moving one with four-momenta $P$ and $Q$ is given by  \cite{F,FG}
\begin{eqnarray}\label{wft}
\Psi_{d,P}^{\rho\omega}({\bf p})
&=&D^{1/2}_{b,~\rho\alpha}(R^W_{L_{P}})
D^{1/2}_{s,~\omega\beta}(R^W_{L_{P}})
\Psi_{d,0}^{\alpha\beta}({\bf p}),\cr\cr
\bar\Psi_{d',Q}^{\lambda\sigma}({\bf q})
&=&\bar\Psi^{\varepsilon\tau}_{d',0}({\bf q})D^{\dag~1/2}_{a,~\varepsilon
\lambda}(R^W_{L_{Q}})D^{\dag~1/2}_{s,~\tau\sigma}(R^W_{L_{Q}}),
\end{eqnarray}
where $R^W$ is the Wigner rotation, $L_{P}$ is the Lorentz boost
from the diquark rest frame to a moving one, and   
the rotation matrix $D^{1/2}(R)$ is defined by
\begin{equation}\label{d12}
{1 \ \ \,0\choose 0 \ \ \,1}D^{1/2}_{b,s}(R^W_{L_{P}})=
S^{-1}({\bf p}_{1,2})S({\bf P})S({\bf p}),
\end{equation}
where
$$
S({\bf
  p})=\sqrt{\frac{\epsilon(p)+m}{2m}}\left(1+\frac{(\bm{\alpha}{\bf
      p})} {\epsilon(p)+m}\right)
$$
is the usual Lorentz transformation matrix of the four-spinor.

Using relations (\ref{pqm}), (\ref{wft}) we can express the matrix
element (\ref{tmm}) in the form of the trace over spinor
indices of both particles. For this aim the following
relations  are useful
\begin{eqnarray}\label{dft}
S_{\alpha\beta}(\Lambda)u^\lambda_\beta({p})
&=&\sum_{\sigma=\pm1/2}u^\sigma_\alpha(\Lambda{ p})
D^{1/2}_{\sigma\lambda}(R^W_{\Lambda p}),\cr
\bar u_\beta^\lambda({p})S^{-1}_{\beta\alpha}(\Lambda)
&=&\sum_{\sigma=\pm1/2}D^{\dag~1/2}_{\lambda\sigma}
(R^W_{\Lambda p})\bar u^\sigma_\alpha(\Lambda {p}).
\end{eqnarray}
Substituting expressions (\ref{gamma1}), (\ref{wft}) in
Eq.~(\ref{tmm}) and using relations (\ref{d12}), (\ref{dft}) we obtain
\begin{eqnarray}\label{mmt}
\langle d'(Q)|J_\mu^W|d(P)\rangle &=&\int\frac{d^3{p}\,d^3{q}}{(2\pi)^3}
\bar\Psi^{\varepsilon\tau}_{d',0}({\bf q})
\bar u_a^\varepsilon(0)\frac{(\hat q_a+m_a)}{\sqrt{2\epsilon_a(q)
(\epsilon_a({q})+m_a)}}S^{-1}(L_{Q})\cr
&&
\times\gamma_\mu(1-\gamma_5)S(L_{P})
\frac{(\hat p_b+m_b)}{\sqrt{2\epsilon_b(p)(\epsilon_b(p)+m_b)}}u^\alpha_b(0)
\bar u^\tau_s(0)\cr
&&\times\frac{(\hat q_s+m_s)}
{\sqrt{2\epsilon_s(q)(\epsilon_s(q)+m_s)}}S^{-1}(L_{Q})
 S(L_{P})\frac{(\hat p_s+m_s)}
{\sqrt{2\epsilon_s(p)(\epsilon_s(p)+m_s)}}\cr
&&\times u_s^\beta(0)\ 
\delta({\bf p}_2-{\bf q}_2)\Psi^{\alpha\beta}_{d,0}({\bf p}),
\end{eqnarray}
where $p_b=(\epsilon_b(p),{\bf p})$, $p_s=(\epsilon_s(p),-{\bf p})$
and  $q_a=(\epsilon_a(q),{\bf q})$, $q_s=(\epsilon_s(q),-{\bf q})$ (we
use the notation $\hat p\equiv p^\mu\gamma_\mu$). 
 The spin wave functions for the scalar and axial vector diquark states
in the heavy quark rest frame read \cite{Lozano,Guo}
\begin{eqnarray}\label{projo}
{\cal H}^S(0)&=&
\frac{1+\gamma^0}{2\sqrt{2}},\cr
{\cal H}^{AV}(0)&=&
\frac{1+\gamma^0}{2\sqrt{2}}\gamma_5\hat\varepsilon,
\end{eqnarray}
where $\varepsilon_\alpha$ is the polarization vector 
of the axial vector diquark.
Using these diquark spin wave functions we get the following expression
\begin{eqnarray}
\langle d'(Q)|J_\mu^W|d(P)\rangle&=&
\int\frac{d^3{p}\,d^3{q}}{(2\pi)^3}\bar\Psi_{d',0}({\bf
  q})\Psi_{d,0}({\bf p}) 
\delta({\bf p}_2-{\bf q}_2)\cr
&&
\times Tr\Biggl\{S^{T~-1}(L_{Q})
\frac{(\hat q_s+m_s)^T}{\sqrt{2\epsilon_s(q)(\epsilon_s(q)+m_c)}}
\ \bar{\cal H}^{S,AV}(0)
\frac{(\hat q_a+m_a)}{\sqrt{\epsilon_a(q)(\epsilon_a(q)+m_a)}}\cr
&&\times
S^{-1}(L_{Q})[\gamma_\mu(1-\gamma_5)]\ S(L_{P})\frac{(\hat p_b+m_b)}
{\sqrt{2\epsilon_b(p)(\epsilon_b(p)+m_b)}}\cr
&&
\times
{\cal H}^{S,AV}(0)
\frac{(\hat p_s+m_s)^T}{\sqrt{2\epsilon_s(p)(\epsilon_s(p)+m_s)}}S^T(L_{P})
\Biggr\},
\end{eqnarray}
where $\Psi_{d,0}({\bf p})$ is the spin-independent part of the
diquark wave function and the superscript $T$ denotes transposing.
After explicit multiplication of the matrices we get the covariant
(see Ref.~\cite{F}) 
expression  for the transition matrix element 
\begin{equation}\label{dd}
{\langle d'(Q)|J_\mu^W|d(P)\rangle}=  {2\sqrt{M_iM_f}}
\int \frac{d^3p\, d^3 q}{(2\pi)^3}Tr\{\bar\Psi_{d'}(Q,q)\gamma_\mu(1
-\gamma_5) \Psi_d(P,p)\}\delta^3({\bf p}_2-{\bf q}_2),
\end{equation} 
where the amplitudes $\Psi_d$ for the scalar ($S$) and axial vector ($AV$)
diquarks ($d$) are
given by
\begin{eqnarray}
  \label{eq:1}
 \Psi_S(P,p)&=&
\sqrt{\frac{\epsilon_b(p)+m_b}{2\epsilon_b(p)}}
\sqrt{\frac{\epsilon_s(p)+m_s}{2\epsilon_s(p)}}\Biggl[\frac{\hat
  v+1}{2\sqrt{2}} +\frac{\hat v-1}{2\sqrt{2}}\frac{{\tilde
    p}^2}{(\epsilon_b(p)+m_b)(\epsilon_s(p)+m_s)} \cr
&&-\left(\frac{\hat v+1}{2\sqrt{2}}\frac1{\epsilon_s(p)+m_s}+
\frac{\hat v-1}{2\sqrt{2}}\frac1{\epsilon_b(p)+m_b}\right)\hat{\tilde
p}\Biggr]\gamma_0\Phi_S(p),
\end{eqnarray}

\begin{eqnarray}
  \label{eq:2}
\Psi_{AV}(P,p,\varepsilon)&=& 
\sqrt{\frac{\epsilon_b(p)+m_b}{2\epsilon_b(p)}}
\sqrt{\frac{\epsilon_s(p)+m_s}{2\epsilon_s(p)}}\Biggl[\frac{\hat
  v+1}{2\sqrt{2}}\hat{\varepsilon}+ \frac{\hat
  v-1}{2\sqrt{2}}\frac{{\tilde
    p}^2}{(\epsilon_b(p)+m_b)(\epsilon_s(p)+m_s)}\hat{
  \varepsilon}   \cr
&&- \frac{\hat v-1}{2\sqrt{2}}\frac{2({\varepsilon}\cdot{\tilde
    p})\hat{\tilde p}}{(\epsilon_b(p)+m_b)(\epsilon_s(p)+m_s)}
+\frac{\hat v+1}{2\sqrt{2}}\frac{\hat{\varepsilon}\hat{\tilde
p}}{\epsilon_s(p)+m_s}\cr
&& -\frac{\hat v-1}{2\sqrt{2}}\frac{\hat{\tilde p}\hat{
\varepsilon} }{\epsilon_b(p)+m_b}\Biggr]\gamma_0\gamma^5\Phi_{AV}(p).
  \end{eqnarray}
Here 
$\Phi_{d}(p)\equiv\Psi_{d,0}({\bf p})/\sqrt{2M_d}$ is the diquark
wave function in the rest frame normalized to unity. The four-vector 
\begin{equation}
\tilde p=L_P(0,{\bf p})=\left(\frac{({\bf p}
{\bf P})}{M},\ {\bf p}+\frac{({\bf p}{\bf P}){\bf
  P}}{M(E+M)}\right)\end{equation}
has the following properties
\begin{eqnarray}\label{eq}
\tilde p^2&=&-{\bf p}^2,\cr
({\varepsilon}\cdot{\tilde p})&=&-(\bm{\varepsilon}{\bf p}),\cr
(\tilde{p}\cdot v)&=&0.
\end{eqnarray}
The presence of $\delta^3({\bf p}_2-{\bf q}_2)$, with momenta ${\bf
  p}_2$ and ${\bf q}_2$ given by Eq.~(\ref{pqm}), in the decay matrix
  element (\ref{dd}) leads to the additional relations 
\begin{eqnarray}
  \label{eq:p}
  \tilde p_\mu&=&\frac{w\epsilon_s(p)-\epsilon_s(q)}{w^2-1}(wv_\mu -
  v'_\mu)=\frac{\sqrt{{\bf p}^2}}{\sqrt{w^2-1}}(w v_\mu-v'_\mu),\\
\label{eq:q}
 \tilde q_\mu&=&\frac{w\epsilon_s(q)-\epsilon_s(p)}{w^2-1}(wv'_\mu -v_\mu)=
 \frac{\sqrt{{\bf q}^2}}{\sqrt{w^2-1}}(wv'_\mu - v_\mu) ,\\
\label{epsq}
\epsilon_s(q)&=&w\epsilon_s(p)-\sqrt{w^2-1}\sqrt{{\bf p}^2},\qquad
{\bf q}^2=\left(\sqrt{w^2-1}\epsilon_s(p)-w\sqrt{{\bf p}^2}\right)^2,\\
\label{epsp}
\epsilon_s(p)&=&w\epsilon_s(q)-\sqrt{w^2-1}\sqrt{{\bf q}^2},\qquad
{\bf p}^2=\left(\sqrt{w^2-1}\epsilon_s(q)-w\sqrt{{\bf q}^2}\right)^2,
\end{eqnarray}
which allow to express either ${\bf q}$ through  ${\bf p}$ or ${\bf
  p}$ through ${\bf q}$. The argument of the
  $\delta$ function can  then  be rewritten  as
\begin{equation}
  \label{eq:df}
  {\bf p}_2-{\bf q}_2={\bf q}-{\bf
  p}-\frac{\epsilon_s(p)+\epsilon_s(q)}{w+1}({\bf v}'-{\bf v}),
\end{equation}
where $w=(v\cdot v')$.

 Calculating traces in Eq.~(\ref{dd}) and using relations
(\ref{eq:p})--(\ref{epsp}) one can see that the spectator quark
contribution factors out in all decay matrix elements. Indeed all
transition matrix elements have a common factor 
\begin{eqnarray}
  \label{eq:sqd}
  \sqrt{\frac{\epsilon_s(p)+m_s}{2\epsilon_s(p)}}
\sqrt{\frac{\epsilon_s(q)+m_s}{2\epsilon_s(q)}}
  &\Biggl[&1-\sqrt{\frac{w-1}{w+1}}\left(
\frac{\sqrt{{\bf p}^2}}{\epsilon_s(p)+m_s}+\frac{\sqrt{{\bf
  q}^2}}{\epsilon_s(q)+m_s}\right)\cr
&& +\frac{\sqrt{{\bf p}^2}
\sqrt{{\bf q}^2}}{[\epsilon_s(q)+m_s][\epsilon_s(p)+m_s]}\Biggr]
= \sqrt{\frac2{w+1}}I_s(p,q).
\end{eqnarray}
If the $\delta$-function is used to express
${\bf q}$ through  ${\bf p}$ or ${\bf p}$ through ${\bf q}$ then
$I_s(p,q)={\cal I}_s(p)$ or $I_s(p,q)={\cal I}_s(q)$ with
\begin{eqnarray}\label{ipq}
{\cal I}_s(p)&=&
\sqrt{\frac{w\epsilon_s(p)-\sqrt{w^2-1}\sqrt{{\bf p}^2}}{\epsilon_s(p)}}
\ \theta\!\left(\sqrt{\epsilon_s(p)-m_s}-\sqrt{\frac{w-1}{w+1}}
\sqrt{\epsilon_s(p)+m_s}\right)\cr
&&+\frac{m_s}{\sqrt{\epsilon_s(p)
[w\epsilon_s(p)-\sqrt{w^2-1}]}}
\ \theta\!\left(\sqrt{\frac{w-1}{w+1}}\sqrt{\epsilon_s(p)+m_s}
-\sqrt{\epsilon_s(p)-m_s}\right).
\end{eqnarray}

The  weak current matrix elements have
the following covariant decomposition

(a) Scalar to scalar diquark transition ($bc\to bc$)
\begin{equation}
  \label{eq:mlss}
\frac{ \langle S_f(v')|J^V_\mu|S_i(v)\rangle}{\sqrt{M_{S_i}M_{S_f}}}
=h_+(w)(v+v')_\mu+h_-(w)(v-v')_\mu,
\end{equation} 

(b) Scalar to axial vector diquark transition ($bc\to cc$)
\begin{equation}
  \label{eq:mlsav}
\frac{\langle AV(v',\varepsilon')|J^V_\mu|S(v)\rangle}{\sqrt{M_{AV}M_S}}
=ih_V(w)\epsilon_{\mu\alpha\beta\gamma} \varepsilon'^{*\alpha} v'^\beta
 v^\gamma,
\end{equation}
\begin{equation}
  \label{eq:mlsaa}
 \frac{ \langle AV(v',\varepsilon')|J^A_\mu|S(v)\rangle}{\sqrt{M_{AV}M_S}}
=h_{A_1}(w)(w+1)\varepsilon'^*_\mu-h_{A_2}(w)(v\cdot\varepsilon'^*)v_\mu
-h_{A_3}(v\cdot\varepsilon'^*)v'_\mu,
\end{equation}

(c) Axial vector to scalar diquark transition ($bb\to bc$)
\begin{equation}
  \label{eq:mlavs}
\frac{\langle S(v')|J^V_\mu| AV(v,\varepsilon)\rangle}{\sqrt{M_{AV}M_S}}
=ih_V(w)\epsilon_{\mu\alpha\beta\gamma} \varepsilon^\alpha v'^\beta 
v^\gamma, 
\end{equation}
\begin{equation}
  \label{eq:mlasa}
\frac{\langle S(v') |J^A_\mu| AV(v,\varepsilon)\rangle}{\sqrt{M_{AV}M_S}}
=h_{A_1}(w)(w+1)\varepsilon_\mu-\tilde h_{A_2}(w)(v'\cdot\varepsilon)v'_\mu
-\tilde h_{A_3}(v\cdot\varepsilon')v_\mu,
\end{equation}

(d) Axial vector to axial vector diquark transition ($bb\to bc$,
$bc\to cc$)
\begin{eqnarray}
  \label{eq:mlava}
\frac{ \langle AV_f(v',\varepsilon'))|J^V_\mu|AV_i(v,\varepsilon)\rangle}
{\sqrt{M_{AV_i}M_{AV_f}}}
&=&-(\varepsilon'^*\cdot\varepsilon)[h_1(w)(v+v')_\mu+h_2(v-v')_\mu]
+h_3(w)(v\cdot\varepsilon'^*)\varepsilon_\mu\cr
&&+
h_4(w)(v'\cdot\varepsilon)\varepsilon'^*_\mu
-(v\cdot\varepsilon'^*)(v'\cdot\varepsilon)[h_5(w)v_\mu+h_6(w)v'_\mu],
\end{eqnarray}
\begin{eqnarray}
  \label{eq:mlaaa}
\frac{ \langle AV_f(v',\varepsilon'))|J^A_\mu|AV_i(v,\varepsilon)\rangle}
{\sqrt{M_{AV_i}M_{AV_f}}}
&=&i\epsilon_{\mu\alpha\beta\gamma}\Bigl\{
\varepsilon^\beta\varepsilon'^{*\gamma} 
[h_7(w)(v+v')^\alpha+h_8(w)(v-v')^\alpha]\cr
&&\qquad\quad +v'^\beta v^\gamma[h_9(w)(v\cdot\varepsilon'^*)
\varepsilon^\alpha+ 
h_{10}(w)(v'\cdot\varepsilon){\varepsilon'^*}^\alpha]\Bigr\}.
\end{eqnarray}
The transition form factors are expressed through the overlap
integrals of the 
diquark wave functions and are given in the Appendix~\ref{ff}. 
These exact expressions for diquark form factors were obtained without
any assumptions about the 
spectator and active quark masses.   

If we consider the spectator quark to be light and then
take the limit of an infinitely heavy active quark mass, $m_{a,b}\to
\infty$, we can explicitly 
obtain heavy quark symmetry relations for the decay matrix
elements of heavy-light diquarks which are analogous to those of
heavy-light mesons \cite{fn,FG}  
\begin{eqnarray*}
h_+(w)&=&h_V(w)=h_{A_1}(w)=h_{A_3}(w)=
\tilde h_{A_3}(w)=h_1(w)=h_3(w)=h_4(w)=h_7(w)=\xi(w),\\
h_-(w)&=&h_{A_2}(w)=\tilde h_{A_2}(w)=h_2(w)=h_5(w)
=h_6(w)=h_8(w)=h_9(w)=h_{10}(w)=0,
\end{eqnarray*}
with the
Isgur-Wise function 
\begin{equation}\label{iwfd}
\xi(w)=\sqrt{\frac2{w+1}} \int\frac{
  d^3p\, d^3q}{(2\pi)^3}\bar\Phi_F({\bf q})\Phi_I({\bf
p}) I_s(p,q)\
\delta^3\left({\bf p}-{\bf q}+\frac{\epsilon_s(p)+\epsilon_s(q)}{w+1}({\bf
  v}' -{\bf v})\right).
\end{equation}
The diquark transition matrix element should be multiplied by a factor
2 if either the initial or final diquark is composed of two identical
heavy quarks.

For the heavy diquark system we can now apply the
$v/c$ expansion.  First we perform the integration over $q$ in the
form factors
(\ref{hph17})--(\ref{ha3t}) and use relations
(\ref{epsq}). Then, applying the nonrelativistic limit, we get the following
expressions for the form factors.

(a) Scalar to scalar diquark transition
\begin{eqnarray}
  \label{eq:ssff}
  h_+(w)&=&F(w),\cr
h_-(w)&=&-(w+1)f(w)F(w),
\end{eqnarray}

(b)  Scalar to axial vector diquark transition
\begin{eqnarray}
  \label{eq:savff}
h_V(w)&=&[1+(w+1)f(w)]F(w),\cr
h_{A_1}(w)&=&h_{A_3}(w)=[1+(w-1)f(w)]F(w),\cr
h_{A_2}(w)&=&-2f(w)F(w),
\end{eqnarray}

(c) Axial vector to scalar diquark transition
\begin{eqnarray}
  \label{eq:avsff}
h_V(w)&=&\tilde h_{A_3}(w)=[1+(w+1)f(w)]F(w),\cr
h_{A_1}(w)&=&[1+(w-1)f(w)]F(w),\cr
\tilde h_{A_2}(w)&=&0,
\end{eqnarray}

\begin{figure}
  \centering
  \includegraphics{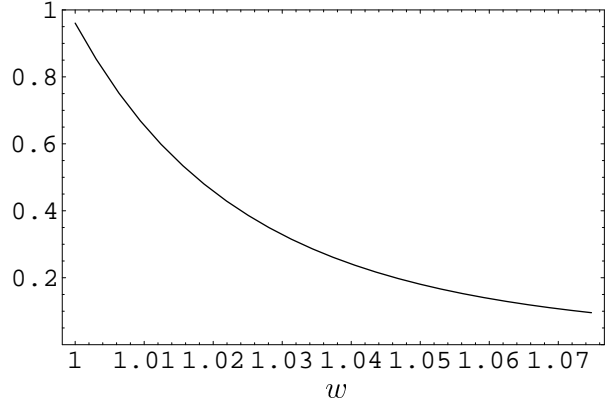}
  \caption{The function $F(w)$ for the $bb\to bc$ quark transition.}
  \label{ffbb}
\end{figure}

\begin{figure}
  \centering
  \includegraphics{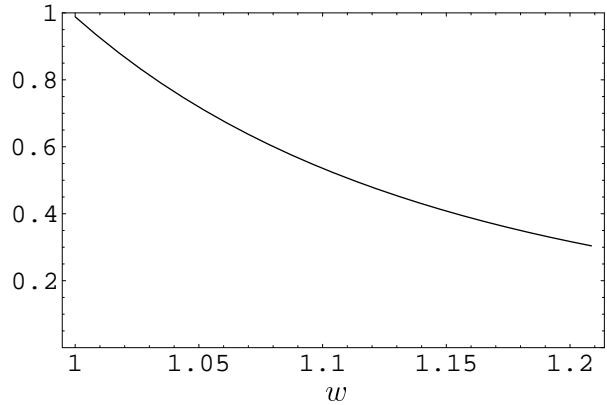}
  \caption{The function $F(w)$ for the $bc\to cc$ quark transition.}
  \label{ffbc}
\end{figure}

(d) Axial vector to axial vector diquark transition
\begin{eqnarray}
  \label{eq:avavff}
h_1(w)&=&h_7(w)=F(w),\cr
h_2(w)&=&h_8(w)=-(w+1)f(w)F(w),\cr
h_3(w)&=&h_4(w)=(1+(w+1)f(w))F(w),\cr
h_5(w)&=&h_9(w)=2f(w)F(w),\cr
h_6(w)&=&h_{10}(w)=0,
\end{eqnarray}
where
\begin{equation}
  \label{eq:ff}
  F(w)=\sqrt{\frac{1}{w(w+1)}}\left(1+
\frac{m_a}{\sqrt{m^2_a+(w^2-1)m_s^2}}\right)^{1/2} \int\frac{{\rm
  d}^3p}{(2\pi)^3}\bar\Phi_F\left({\bf p}+\frac{2m_s}{w+1}({\bf v}'
-{\bf v})\right)\Phi_I({\bf p})
\end{equation}
and
\begin{equation}
  \label{eq:fw}
  f(w)=\frac{m_s}{\sqrt{m^2_a+(w^2-1)m_s^2}+m_a}.
\end{equation}
The appearance of the terms proportional to the function $f(w)$ is the
result of the account of 
the spectator quark recoil. Their contribution is important and
distinguishes our approach from previous considerations
\cite{WS,Lozano}. We plot the function $F(w)$ for $bb\to bc$ and
$bc\to cc$ diquark transitions in Figs.~\ref{ffbb},~\ref{ffbc}. 


\section{Doubly heavy baryon transitions}
\label{dhbt}

The second step in studying weak transitions of doubly
heavy baryons is the inclusion of the spectator light quark in
the consideration. We carry out all further calculations in the limit
of an
infinitely heavy diquark mass, $M_d\to\infty$, treating the light quark
relativistically.  The transition matrix element between doubly heavy
baryon states in the quark-diquark approximation (see
Fig.~\ref{mdiag} and \ref{diag}) is given by [cf. Eqs.~(\ref{tmm}) and
(\ref{gamma1})] 
\begin{equation}\label{dd1}
{\langle B'(Q)|J_\mu^W|B(P)\rangle}=  {2\sqrt{M_IM_F}}
\int \frac{d^3p\, d^3 q}{(2\pi)^3}\bar \Psi_{B',Q}({\bf q})
\langle d'(Q)|J_\mu^W|d(P)\rangle
 \Psi_{B,P}({\bf p})\delta^3({\bf p}_q-{\bf q}_q),
\end{equation} 
where $\Psi_{B,P}(p)$ is the doubly heavy baryon wave function; ${\bf
  p}$ and ${\bf q}$ are the relative quark-diquark momenta, ${\bf
  p}_q$ and ${\bf q}_q$ are light quark momenta expressed in the form
  similar to (\ref{pqm}).
 
The baryon ground-state wave function $\Psi_{B,P}(p)$ is a product of
the spin-independent part $\Psi_B(p)$ satisfying the related
quasipotential equation (\ref{quas}) and the spin part $U_B(v)$
\begin{equation}
  \label{eq:psib}
  \Psi_{B,P}({\bf p})=\Psi_B({\bf p})U_B(v).
\end{equation}
The baryon spin wave function is constructed from the Dirac spinor
$u_q(v)$ of the light spectator quark and the diquark wave
function. The ground state spin 1/2 baryons can 
contain either the scalar  or axial vector
diquark. The former baryon is denoted by $\Xi'_{QQ'}$ and
the latter one by $\Xi_{QQ'}$. The ground state spin 3/2 baryon can be
formed only from axial vector diquark and is denoted by $\Xi^*_{QQ'}$. To
obtain the corresponding baryon spin-states we use in the baryon
matrix elements the following replacements
\begin{eqnarray}
\label{sdb}
u_q(v) &\to & U_{\Xi'_{QQ'}}(v),\cr
[\varepsilon_\mu(v)u_q(v)]_{\rm spin~1/2}&\to
&\frac{i}{\sqrt{3}}(\gamma_\mu+v_\mu) 
\gamma_5 U_{\Xi_{QQ'}}(v),\cr
[\varepsilon^\mu(v)u_q(v)]_{\rm spin~3/2}&\to & {U^\mu_{\Xi^*_{QQ'}}(v)},
\end{eqnarray}  
where baryon spinor wave functions are normalized by $U^*_B U_B=1$
($B=\Xi',\Xi$) and the Rarita-Schwinger wave functions
are normalized by $U_{\Xi^*}^{*\mu} U_{\Xi^*\mu}=-1$.

 Then the decay amplitudes of doubly heavy baryons in the infinitely heavy
diquark limit are given by the following expressions.

(a) $\Xi'_{QQ_s}\to\Xi'_{Q'Q_s}$ transition
\begin{equation}
  \label{eq:ltol}
\frac{\langle\Xi'_{Q'Q_s}(v')|J^W_\mu|\Xi'_{QQ_s}(v)\rangle}{2\sqrt{M_IM_F}}
=[ h_+(w)(v+v')_\mu+h_-(w)(v-v')_\mu]\bar U_{\Xi'_{Q'Q_s}}(v')
U_{\Xi'_{QQ_s}}(v)\eta(w),
\end{equation}

(b) $\Xi'_{QQ_s}\to\Xi_{Q'Q_s}$ and $\Xi_{QQ_s}\to\Xi'_{Q'Q_s}$ transitions
\begin{eqnarray}
  \label{eq:ltos}
\frac{\langle\Xi_{Q'Q_s}(v')|J^W_\mu|\Xi'_{QQ_s}(v)\rangle}{2\sqrt{M_IM_F}}
&=&\frac{i}{\sqrt{3}}[i h_V(w)\epsilon_{\mu\alpha\beta\gamma} v'^\beta
v^\gamma 
-g_{\mu\alpha}h_{A_1}(w+1)+v_\mu v_\alpha h_{A_2}(w)\cr
&&
+v'_\mu v_\alpha h_{A_3}(w)]\bar U_{\Xi_{Q'Q_s}}(v')\gamma_5(\gamma^\alpha
+v'^\alpha)  U_{\Xi'_{QQ_s}}(v)\eta(w),\\
\frac{\langle\Xi'_{Q'Q_s}(v')|J^W_\mu|\Xi_{QQ_s}(v)\rangle}{2\sqrt{M_IM_F}}
&=&\frac{i}{\sqrt{3}}[i h_V(w)\epsilon_{\mu\alpha\beta\gamma}v'^\beta
v^\gamma 
-g_{\mu\alpha} h_{A_1}(w+1)+v'_\mu v'_\alpha\tilde h_{A_2}(w)\cr
&&
+v_\mu v'_\alpha\tilde h_{A_3}(w)]\bar U_{\Xi'_{Q'Q_s}}(v')(\gamma^\alpha
+v^\alpha) \gamma_5 U_{\Xi_{QQ_s}}(v)\eta(w),
\end{eqnarray}

(c) $\Xi'_{QQ_s}\to\Xi^*_{Q'Q_s}$ and $\Xi^*_{QQ_s}\to\Xi'_{Q'Q_s}$
transitions 
\begin{eqnarray}
  \label{eq:ltos3}
 \frac{\langle\Xi^*_{Q'Q_s}(v')|J^W_\mu|\Xi'_{QQ_s}(v)\rangle}
 {2\sqrt{M_IM_F}} 
&=&[i h_V(w)\epsilon_{\mu\alpha\beta\gamma}v'^\beta v^\gamma
-g_{\mu\alpha}h_{A_1}(w+1)+v_\mu v_\alpha h_{A_2}(w)\cr
&&
+v'_\mu v_\alpha h_{A_3}(w)]\bar U_{\Xi^*_{Q'Q_s}}^\alpha(v')
U_{\Xi'_{QQ_s}}(v)\eta(w),\\
\frac{\langle\Xi'_{Q'Q_s}(v')|J^W_\mu|\Xi^*_{QQ_s}(v)\rangle}
 {2\sqrt{M_IM_F}} 
&=&[i h_V(w)\epsilon_{\mu\alpha\beta\gamma}v'^\beta v^\gamma
-g_{\mu\alpha}h_{A_1}(w+1)+v'_\mu v'_\alpha\tilde h_{A_2}(w)\cr
&&
+v_\mu v'_\alpha\tilde h_{A_3}(w)]\bar U_{\Xi'_{Q'Q_s}}(v')
U_{\Xi^*_{QQ_s}}^\alpha(v)\eta(w),  
\end{eqnarray}

(d)  $\Xi_{QQ_s}\to\Xi_{Q'Q_s}$ transition
\begin{eqnarray}
  \label{eq:s1tos1}
\frac{\langle\Xi_{Q'Q_s}(v')|J^W_\mu|\Xi_{QQ_s}(v)\rangle}{2\sqrt{M_IM_F}}
&=&-\frac13\biggl\{g_{\rho\lambda}[h_1(w)(v+v')_\mu+h_2(w)(v-v')_\mu]
-h_3(w)g_{\mu\rho}v_\lambda\cr
&&-h_4(w)g_{\mu\lambda}v'_\rho +v'_\rho
v_\lambda [h_5(w)v_\mu+h_6(w)v'_\mu]\cr
&&
+i\epsilon_{\mu\alpha\beta\gamma}\Bigl
(g^\beta_\rho g^\gamma_\lambda[h_7(w)(v+v')^\alpha+h_8(w)(v-v')^\alpha]\cr
&&
+v'^\beta v^\gamma[h_9(w)g^\alpha_\rho v_\lambda+h_{10}(w)
g^\alpha_\lambda v'_\rho]\Bigr)\biggr\}\bar U_{\Xi_{Q'Q_s}}(v')
\gamma_5(\gamma^\lambda+v'^\lambda)\cr
&&\times (\gamma^\rho
+v^\rho) \gamma_5 U_{\Xi_{QQ_s}}(v)\eta(w),  
\end{eqnarray}

(e) $\Xi_{QQ_s}\to\Xi^*_{Q'Q_s}$ and $\Xi^*_{QQ_s}\to\Xi_{Q'Q_s}$
transitions 
\begin{eqnarray}
  \label{eq:s1tos3}
\frac{\langle\Xi^*_{Q'Q_s}(v')|J^W_\mu|\Xi_{QQ_s}(v)\rangle}{2\sqrt{M_IM_F}}
&=&-\frac{i}{\sqrt3}\biggl\{g_{\rho\lambda}[h_1(w)(v+v')_\mu
+h_2(w)(v-v')_\mu] 
-h_3(w)g_{\mu\rho}v_\lambda\cr
&&-h_4(w)g_{\mu\lambda}v'_\rho +v'_\rho
v_\lambda [h_5(w)v_\mu+h_6(w)v'_\mu]\cr
&&
+i\epsilon_{\mu\alpha\beta\gamma}\Bigl
(g^\beta_\rho g^\gamma_\lambda[h_7(w)(v+v')^\alpha+h_8(w)(v-v')^\alpha]\cr
&&
+v'^\beta v^\gamma[h_9(w)g^\alpha_\rho v_\lambda+h_{10}(w)
g^\alpha_\lambda v'_\rho]\Bigr)\biggr\}\bar
U_{\Xi^*_{Q'Q_s}(v')}^\lambda(v') 
\cr
&&\times (\gamma^\rho
+v^\rho) \gamma_5 U_{\Xi_{QQ_s}}(v)\eta(w),\\  
\frac{\langle\Xi_{Q'Q_s}(v')|J^W_\mu|\Xi^*_{QQ_s}(v)\rangle}{2\sqrt{M_IM_F}}
&=&-\frac{i}{\sqrt3}\biggl\{g_{\rho\lambda}[h_1(w)(v+v')_\mu
+h_2(w)(v-v')_\mu] 
-h_3(w)g_{\mu\rho}v_\lambda\cr
&&-h_4(w)g_{\mu\lambda}v'_\rho +v'_\rho
v_\lambda [h_5(w)v_\mu+h_6(w)v'_\mu]\cr
&&
+i\epsilon_{\mu\alpha\beta\gamma}\Bigl
(g^\beta_\rho g^\gamma_\lambda[h_7(w)(v+v')^\alpha+h_8(w)(v-v')^\alpha]\cr
&&
+v'^\beta v^\gamma[h_9(w)g^\alpha_\rho v_\lambda+h_{10}(w)
g^\alpha_\lambda v'_\rho]\Bigr)\biggr\}\bar U_{\Xi_{Q'Q_s}}(v')\cr
&&\times
\gamma_5(\gamma^\lambda+v'^\lambda) U_{\Xi^*_{QQ_s}}^\rho(v)\eta(w),
\end{eqnarray}

\begin{figure}
  \centering
  \includegraphics{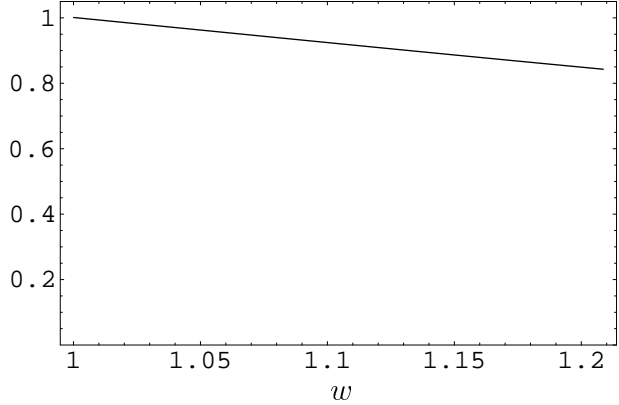}
  \caption{The Isgur-Wise function $\eta(w)$ of the light quark --
  heavy diquark bound system.}
  \label{fig:eta}
\end{figure}

(f)  $\Xi^*_{QQ_s}\to\Xi^*_{Q'Q_s}$ transition
\begin{eqnarray}
  \label{eq:s3tos3}
\frac{\langle\Xi^*_{Q'Q_s}(v')|J^W_\mu|\Xi^*_{QQ_s}(v)\rangle}
{2\sqrt{M_IM_F}} 
&=&-\biggl\{g_{\rho\lambda}[h_1(w)(v+v')_\mu+h_2(w)(v-v')_\mu]
-h_3(w)g_{\mu\rho}v_\lambda\cr
&&-h_4(w)g_{\mu\lambda}v'_\rho +v'_\rho
v_\lambda [h_5(w)v_\mu+h_6(w)v'_\mu]\cr
&&
+i\epsilon_{\mu\alpha\beta\gamma}\Bigl
(g^\beta_\rho g^\gamma_\lambda[h_7(w)(v+v')^\alpha+h_8(w)(v-v')^\alpha]\cr
&&
+v'^\beta v^\gamma[h_9(w)g^\alpha_\rho v_\lambda+h_{10}(w)
g^\alpha_\lambda v'_\rho]\Bigr)\biggr\}\bar
U_{\Xi^*_{Q'Q_s}}^\lambda (v')\cr
&&\times
 U_{\Xi^*_{QQ_s}}^\rho(v)\eta(w).  
\end{eqnarray}
Here $\eta(w)$ is the heavy
diquark -- light quark Isgur-Wise function which is determined by the
dynamics of the light spectator quark $q$ and is calculated similarly to
Eq.~(\ref{iwfd})
 \begin{equation}
  \label{eq:eta}
\eta(w)=\sqrt{\frac2{w+1}} \int\frac{d^3p\, 
d^3q}{(2\pi)^3}\bar\Psi_B({\bf q})\Psi_B({\bf
p}) I_q(p,q)\
\delta^3\!\left({\bf p}-{\bf q}+\frac{\epsilon_q(p)+\epsilon_q(q)}{w+1}({\bf
  v}' -{\bf v})\right).
  \end{equation}
We plot the Isgur-Wise function $\eta(w)$ in Fig.~\ref{fig:eta}. In
the nonrelativistic limit for heavy quarks the diquark form factors
$h_i(w)$ obey relations (\ref{eq:ssff})--(\ref{eq:avavff}). In this
limit the baryon transition matrix elements contain the common factor
$F(w)\eta(w)$ (cf. \cite{WS}).   

\section{Semileptonic decay rates of doubly heavy baryons}
\label{sdr}

The exclusive differential rate of the doubly heavy
baryon semileptonic decay
$B\to B'e\bar\nu$ can be written in the form
\begin{equation}
\frac{d\Gamma}{dw}=\frac{G_F^2|V_{bc}|^2M_F^3\sqrt{w^2-1}}
{48\pi^3}\Omega(w),
\end{equation}
where $w=(v\cdot v')$=$(M_I^2+M_F^2-k^2)/(2M_IM_F)$, $k=P-Q$
and the  function
$\Omega(w)$ is the contraction of the hadronic transition matrix
elements and the leptonic tensor. For
the massless leptons the differential decay rates of the transitions
$\Xi'_{QQ_s}\to \Xi_{Q'Q_s}$ and $\Xi_{QQ_s}\to \Xi'_{Q'Q_s}$  are as
follows 
\begin{eqnarray}
  \label{eq:dgrls1}
  \frac{{d}\Gamma}{{d}w}(\Xi'_{QQ_s}\to\Xi_{Q'Q_s})
&=&\frac{G_F^2|V_{QQ'}|^2}{72\pi^3}(w^2-1)^{1/2}(w+1)^3M_F^3
\Biggl\{2(M_I^2+M_F^2-2M_IM_Fw)\cr
&&\times\left[h_{A_1}^2(w)
+\frac{w-1}{w+1}h_V^2(w)\right]+\biggl[(M_F-M_Iw)\cr
&&\times h_{A_1}(w)+(w-1)
\left(M_Fh_{A_2}(w)+M_Ih_{A_3}(w)
\right)\biggr]^2\Biggr\}\eta^2(w),
\end{eqnarray}
\begin{eqnarray}
  \label{eq:dgrs1l}
  \frac{{d}\Gamma}{{d}w}(\Xi_{QQ_s}\to\Xi'_{Q'Q_s})
&=&\frac{M_F^3}{M_I^3} \frac{{d}\Gamma}{{d}w}(\Xi'_{QQ_s}\to\Xi_{Q'Q_s},
M_I\leftrightarrow M_F, h_{A_2}\to\tilde h_{A_2},h_{A_3}\to\tilde
h_{A_3}), \quad
\end{eqnarray}
The differential decay rates for other transitions are given in the
Appendix~\ref{ddr}.  

The semileptonic decay rates of doubly heavy baryons are calculated in
the nonrelativistic limit for heavy quarks and presented in
Tables~\ref{dhbdr} and  \ref{dhbdrs}.

\begin{table}
\caption{\label{dhbdr}Semileptonic decay rates of doubly heavy baryons
  $\Xi_{bb}$ and $\Xi_{bc}$ 
(in $\times 10^{-14}$ Gev).}
\begin{ruledtabular}
\begin{tabular}{cccccc}
Decay & our &Ref.\cite{Guo}&Ref.\cite{Lozano} & Ref.\cite{AO} &
Ref.\cite{Ivanov} \\  \hline
$\Xi_{bb}\to\Xi'_{bc}$ & $1.64$ & $4.28$ &  &    &   \\   
$\Xi_{bb}\to\Xi_{bc}$ & $3.26$ & $28.5$ &  & $8.99$ &    \\   
$\Xi_{bb}\to\Xi^*_{bc}$ & $1.05$ & $27.2$ &  & $2.70$ & \\
$\Xi^*_{bb}\to\Xi'_{bc}$ & $1.63$ & $8.57$ &  &  &  \\ 
$\Xi^*_{bb}\to\Xi_{bc}$ & $0.55$ & $52.0$ &   &   & \\  
$\Xi^*_{bb}\to\Xi^*_{bc}$ & $3.83$ & $12.9$ &   &   & \\   
$\Xi'_{bc}\to\Xi_{cc}$ & $1.76$ & $7.76$    &   &  & \\  
$\Xi'_{bc}\to\Xi^*_{cc}$ & $3.40$ & $28.8$    &   &  & \\
$\Xi_{bc}\to\Xi_{cc}$ & $4.59$ & $8.93$ & $4.0$ & $8.87$& $0.8$ \\
$\Xi_{bc}\to\Xi^*_{cc}$ & $1.43$ & $14.1$ &   $1.2$ & $2.66$ &     \\ 
$\Xi^*_{bc}\to\Xi_{cc}$ & $0.75$ & $27.5$ &   &  & \\     
$\Xi^*_{bc}\to\Xi^*_{cc}$ & $5.37$ & $17.2$ &   &  &  \\  
\end{tabular}
\end{ruledtabular}
\end{table}

\begin{table}
\caption{\label{dhbdrs}Semileptonic decay rates of doubly heavy baryons
  $\Omega_{bb}$ and $\Omega_{bc}$ 
(in $\times 10^{-14}$ Gev).}
\begin{ruledtabular}
\begin{tabular}{cccc}
Decay & $\Gamma$ & Decay  & $\Gamma$ \\ \hline
$\Omega_{bb}\to\Omega'_{bc}$ & 1.66 & $\Omega'_{bc}\to\Omega_{cc}$& 1.90\\
$\Omega_{bb}\to\Omega_{bc}$ & 3.40 &  $\Omega'_{bc}\to\Omega^*_{cc}$ &3.66\\
$\Omega_{bb}\to\Omega^*_{bc}$ & 1.10 &  $\Omega_{bc}\to\Omega_{cc}$ &4.95 \\
$\Omega^*_{bb}\to\Omega'_{bc}$ & 1.70 & $\Omega_{bc}\to\Omega^*_{cc}$&1.48\\
$\Omega^*_{bb}\to\Omega_{bc}$ & 0.57 & $\Omega^*_{bc}\to\Omega_{cc}$&0.80\\
$\Omega^*_{bb}\to\Omega^*_{bc}$ & 3.99 &
$\Omega^*_{bc}\to\Omega^*_{cc}$ & 5.76\\ 
\end{tabular}
\end{ruledtabular}
\end{table}

\section{Conclusions} 
\label{conc}
In this paper we calculated the semileptonic decay rates of doubly
heavy baryons in the quark-diquark approximation. The weak transition
matrix elements between heavy diquark states were calculated with the
self-consistent account of the spectator quark
recoil. It was shown that recoil effects lead to the
additional contributions to the transition matrix elements.  Such
terms were missed in the previous quark model calculations. If we
neglect these recoil contributions, the previously obtained expressions
\cite{WS,Lozano} for heavy 
diquark transition matrix elements are reproduced. It was found that
these recoil terms which are proportional to the ratio of the heavy
spectator to the final active quark mass [see
Eqs. (\ref{eq:ssff})--(\ref{eq:fw})] give important contributions to
transition matrix elements of doubly heavy diquarks even in the
nonrelativistic limit. In this limit these weak transition matrix
elements are proportional to the function $F(w)$ (\ref{eq:ff}) which
is expressed through the overlap integral of the heavy diquark wave
functions. The function $F(w)$ falls off rather rapidly, especially for
$bb\to bc$ diquark transition where the spectator quark is the $b$
quark (see Figs~\ref{ffbb}, \ref{ffbc}). Such a decrease is the
consequence of the large mass of the spectator quark and 
high recoil momenta ($q_{\rm max}\approx m_b-m_c\sim 3.33$~GeV)
transfered.    

We calculated the doubly heavy baryon transition matrix elements in
the heavy diquark limit. The expressions for transition amplitudes and
decay rates were obtained for the most general parameterization of
the diquark transition matrix elements. The Isgur-Wise function
$\eta(w)$ (\ref{eq:eta}) for the light 
quark -- heavy diquark bound system was determined. This function is
very similar to the Isgur-Wise function of the heavy-light meson in our
model \cite{FG} as it is required by the heavy quark symmetry. In the
heavy quark limit the baryon transition matrix elements contain the
common factor which is the product of the diquark form factor $F(w)$
and the Isgur-Wise function $\eta(w)$. 

Our results for the semileptonic decay rates of doubly heavy
baryons $\Xi_{bb}$ and $\Xi_{bc}$ are compared with previous
predictions in Table~\ref{dhbdr}. It is seen from this Table that
results of different approaches differ substantially. Most of previous
papers \cite{Ivanov,Lozano,AO} give their predictions only for
selected decay modes. Their values agree with our in the order of magnitude.
Our predictions are smaller than the QCD sum rule
results \cite{AO} by a factor of $\sim2$.  This can be a result of our
treatment of the heavy spectator quark recoil in the heavy diquark. On
the other hand the authors of Ref.~\cite{Guo}, where the
Bethe-Salpeter equation is used, give more decay
channels. Their results are substantially higher than ours, for some
decays the difference reaches almost two orders of magnitude which
seems quite strange. E.g., for the
sum of the semileptonic decays $\Xi_{bb}\to\Xi_{bc}^{(\prime,*)}$
Ref.~\cite{Guo} predicts $\sim 6\times 10^{-13}$ which almost saturates
the estimate of the total decay rate $\Gamma_{\Xi_{bb}}^{\rm
  toatal}\sim (8.3\pm 
0.3)\times 10^{-13}$ \cite{UFN} and thus is unlikely.    

All considerations in the present paper were done in the heavy quark
limit. The calculations of the semileptonic decay rates of heavy mesons
indicate that the $1/m_Q$ corrections to the heavy quark limit
(especially $1/m_c$) are important \cite{MN,fn,FG,FGM}. Thus they can
give sizeable 
contributions also to the semileptonic decay rates of doubly heavy
baryons. Their account will result in the more complicated structure
of the baryon matrix elements. We plan to study these corrections in
future.

\acknowledgments
We are grateful to M. Ivanov, V. Kiselev, A. Likhoded, V. Luybovitskij,
M. M\"uller-Preussker, A. Onishchenko and V. Savrin for useful
discussions. Two of us (R.N.F and V.O.G.) were supported in part by the 
{\it Deutsche Forschungsgemeinschaft} under contract Eb 139/2-2.
A.P.M. was supported in part by the German Academic
Exchange Service (DAAD).

\appendix

\section{Form factors of the diquark transitions}
\label{ff}
 
\begin{eqnarray}\label{hph17}
h_+(w)=h_1(w)=h_7(w)&=&\sqrt{\frac2{w+1}}\int\frac{
  d^3p\, d^3q}{(2\pi)^3}\bar\Phi_{d'}({\bf q})
\sqrt{\frac{\epsilon_a(q)+m_a}{2\epsilon_a(q)}}
\sqrt{\frac{\epsilon_b(p)+m_b}{2\epsilon_b(p)}}\cr
&&\times\left(1-\frac{\sqrt{{\bf p}^2}\sqrt{{\bf q}^2}}
{[\epsilon_a(q)+m_a][\epsilon_b(p)+m_b]}\right)\Phi_{d}({\bf p})\cr
&&\times I_s(p,q)\
\delta^3\!\left({\bf p}-{\bf q}+\frac{\epsilon_s(p)+\epsilon_s(q)}{w+1}({\bf
  v}' -{\bf v})\right),
\end{eqnarray}
\begin{eqnarray}\label{hmh28}
h_-(w)=h_2(w)=h_8(w)&=&\sqrt{\frac2{w+1}}\int\frac{
  d^3p\, d^3q}{(2\pi)^3}\bar\Phi_{d'}({\bf q})
\sqrt{\frac{\epsilon_a(q)+m_a}{2\epsilon_a(q)}}
\sqrt{\frac{\epsilon_b(p)+m_b}{2\epsilon_b(p)}}\cr
&&\times
\sqrt{\frac{w+1}{w-1}}\left(\frac{\sqrt{{\bf q}^2}}
{\epsilon_a(q)+m_a}-\frac{\sqrt{{\bf p}^2}}
{\epsilon_b(p)+m_b}\right)\Phi_{d}({\bf p})\cr
&&\times I_s(p,q)\
\delta^3\!\left({\bf p}-{\bf q}+\frac{\epsilon_s(p)+\epsilon_s(q)}{w+1}({\bf
  v}' -{\bf v})\right),
\end{eqnarray}
\begin{eqnarray}
  \label{hvh34}
h_V(w)=h_3(w)=h_4(w)
&=&
\sqrt{\frac2{w+1}}\int\frac{d^3p\, d^3q}{(2\pi)^3}\bar\Phi_{d'}({\bf q})
\sqrt{\frac{\epsilon_a(q)+m_a}{2\epsilon_a(q)}}
\sqrt{\frac{\epsilon_b(p)+m_b}{2\epsilon_b(p)}}\cr
&&\times
\Biggl[1 +\sqrt{\frac{w+1}{w-1}}\left(\frac{\sqrt{{\bf q}^2}}
{\epsilon_a(q)+m_a}+\frac{\sqrt{{\bf p}^2}}
{\epsilon_b(p)+m_b}\right)\cr
&& \qquad +\frac{\sqrt{{\bf p}^2}\sqrt{{\bf q}^2}}
{[\epsilon_a(q)+m_a][\epsilon_b(p)+m_b]}\Biggr] \Phi_{d}({\bf p})
\cr
&&\times
I_s(p,q)\
\delta^3\!\left({\bf p}-{\bf q}+\frac{\epsilon_s(p)+\epsilon_s(q)}{w+1}({\bf
  v}' -{\bf v})\right),
\end{eqnarray}
\begin{eqnarray}
  \label{ha1}
 h_{A_1}(w)&=&
\sqrt{\frac2{w+1}}\int\frac{
  d^3p\, d^3q}{(2\pi)^3}\bar\Phi_{d'}({\bf q})
\sqrt{\frac{\epsilon_a(q)+m_a}{2\epsilon_a(q)}}
\sqrt{\frac{\epsilon_b(p)+m_b}{2\epsilon_b(p)}}\cr
&&\times
\Biggl[1 +\sqrt{\frac{w-1}{w+1}}\left(\frac{\sqrt{{\bf q}^2}}
{\epsilon_a(q)+m_a}+\frac{\sqrt{{\bf p}^2}}
{\epsilon_b(p)+m_b}\right) +\frac{\sqrt{{\bf p}^2}\sqrt{{\bf q}^2}}
{[\epsilon_a(q)+m_a][\epsilon_b(p)+m_b]}\Biggr]\cr
&&\times\Phi_{d}({\bf p})
I_s(p,q)\
\delta^3\!\left({\bf p}-{\bf q}+\frac{\epsilon_s(p)+\epsilon_s(q)}{w+1}({\bf
  v}' -{\bf v})\right),
\end{eqnarray}

\begin{eqnarray}
  \label{ha259}
 h_{A_2}(w)= h_5(w)= h_9(w)&=&
\sqrt{\frac2{w+1}}\int\frac{
  d^3p\, d^3q}{(2\pi)^3}\bar\Phi_{d'}({\bf q})
\sqrt{\frac{\epsilon_a(q)+m_a}{2\epsilon_a(q)}}
\sqrt{\frac{\epsilon_b(p)+m_b}{2\epsilon_b(p)}}\cr
&&\times
\frac{2\sqrt{{\bf q}^2}}{\sqrt{w^2-1}[\epsilon_a(q)+m_a]}
\left(1+\sqrt{\frac{w+1}{w-1}}\frac{\sqrt{{\bf p}^2}}
{\epsilon_b(p)+m_b}\right)\Phi_{d}({\bf p})\cr
&&\times
I_s(p,q)\
\delta^3\!\left({\bf p}-{\bf q}+\frac{\epsilon_s(p)+\epsilon_s(q)}{w+1}({\bf
  v}' -{\bf v})\right),
\end{eqnarray}
\begin{eqnarray}
  \label{ha2610}
\tilde h_{A_2}(w)= h_6(w)= h_{10}(w)&=&
\sqrt{\frac2{w+1}}\int\frac{
  d^3p\, d^3q}{(2\pi)^3}\bar\Phi_{d'}({\bf q})
\sqrt{\frac{\epsilon_a(q)+m_a}{2\epsilon_a(q)}}
\sqrt{\frac{\epsilon_b(p)+m_b}{2\epsilon_b(p)}}\cr
&&\times
\frac{2\sqrt{{\bf p}^2}}{\sqrt{w^2-1}[\epsilon_b(p)+m_b]}
\left(1+\sqrt{\frac{w+1}{w-1}}\frac{\sqrt{{\bf q}^2}}
{\epsilon_a(q)+m_a}\right)\Phi_{d}({\bf p})\cr
&&\times
I_s(p,q)\
\delta^3\!\left({\bf p}-{\bf q}+\frac{\epsilon_s(p)+\epsilon_s(q)}{w+1}({\bf
  v}' -{\bf v})\right),
\end{eqnarray}
\begin{eqnarray}
  \label{ha3}
h_{A_3}(w)&=&
\sqrt{\frac2{w+1}}\int\frac{
  d^3p\, d^3q}{(2\pi)^3}\bar\Phi_{d'}({\bf q})
\sqrt{\frac{\epsilon_a(q)+m_a}{2\epsilon_a(q)}}
\sqrt{\frac{\epsilon_b(p)+m_b}{2\epsilon_b(p)}}\cr
&&\times
\Biggl[1 +\sqrt{\frac{w+1}{w-1}}\left(\frac{w-1}{w+1}\frac{\sqrt{{\bf q}^2}}
{\epsilon_a(q)+m_a}+\frac{\sqrt{{\bf p}^2}}
{\epsilon_b(p)+m_b}\right)\cr
&& +\frac{w+1}{w-1}\frac{\sqrt{{\bf p}^2}\sqrt{{\bf q}^2}}
{[\epsilon_a(q)+m_a][\epsilon_b(p)+m_b]}\Biggr]\Phi_{d}({\bf p})
\cr
&&\times
I_s(p,q)\
\delta^3\!\left({\bf p}-{\bf q}+\frac{\epsilon_s(p)+\epsilon_s(q)}{w+1}({\bf
  v}' -{\bf v})\right),
\end{eqnarray}
\begin{eqnarray}
  \label{ha3t}
\tilde h_{A_3}(w)&=&
\sqrt{\frac2{w+1}}\int\frac{
  d^3p\, d^3q}{(2\pi)^3}\bar\Phi_{d'}({\bf q})
\sqrt{\frac{\epsilon_a(q)+m_a}{2\epsilon_a(q)}}
\sqrt{\frac{\epsilon_b(p)+m_b}{2\epsilon_b(p)}}\cr
&&\times
\Biggl[1 +\sqrt{\frac{w+1}{w-1}}\left(\frac{\sqrt{{\bf q}^2}}
{\epsilon_a(q)+m_a}+\frac{w-1}{w+1}\frac{\sqrt{{\bf p}^2}}
{\epsilon_b(p)+m_b}\right)\cr
&& +\frac{w+1}{w-1}\frac{\sqrt{{\bf p}^2}\sqrt{{\bf q}^2}}
{[\epsilon_a(q)+m_a][\epsilon_b(p)+m_b]}\Biggr]\Phi_{d}({\bf p})
\cr
&&\times
I_s(p,q)\
\delta^3\!\left({\bf p}-{\bf q}+\frac{\epsilon_s(p)+\epsilon_s(q)}{w+1}({\bf
  v}' -{\bf v})\right).
\end{eqnarray}

\section{Differential decay rates of doubly heavy baryons}
\label{ddr}
\begin{eqnarray}
  \label{eq:dgrll}
  \frac{{d}\Gamma}{{d}w}(\Xi'_{QQ_s}\to\Xi'_{Q'Q_s})
&=&\frac{G_F^2|V_{QQ'}|^2}{24\pi^3}(w^2-1)^{3/2}(w+1)M_F^3(M_I+M_F)^2\cr
&&\times
\left[h_+(w)-\frac{M_I-M_F}{M_I+M_F}h_-(w)\right]^2\eta^2(w),
\end{eqnarray}

\begin{eqnarray}
  \label{eq:dgrls3}
  \frac{{d}\Gamma}{{d}w}(\Xi'_{QQ_s}\to\Xi^*_{Q'Q_s})
&=&\frac{G_F^2|V_{QQ'}|^2}{36\pi^3}(w^2-1)^{1/2}(w+1)^3M_F^3\cr
&&\times
\Biggl\{(M_I^2+M_F^2-2M_IM_Fw)\left[3h_{A_1}^2(w)
+2\frac{w-1}{w+1}h_V^2(w)\right]\cr
&&+(w-1)\biggl[M_I^2(w+1)
h_{A_1}^2(w)+(w-1)\left(M_Fh_{A_2}(w)+M_Ih_{A_3}(w) 
\right)^2\cr
&&+2(M_F-M_Iw)h_{A_1}(w)\left(M_Fh_{A_2}(w)+M_Ih_{A_3}(w)
\right)\biggr]
\Biggr\}\eta^2(w),
\end{eqnarray}
\begin{eqnarray}
  \label{eq:dgrs3l}
  \frac{{d}\Gamma}{{d}w}(\Xi^*_{QQ_s}\to\Xi'_{Q'Q_s})
&=&\frac12\frac{M_F^3}{M_I^3} \frac{{d}\Gamma}{{d}w}
(\Xi'_{QQ_s}\to\Xi^*_{Q'Q_s},
M_I\leftrightarrow M_F, h_{A_2}\to\tilde h_{A_2},h_{A_3}\to\tilde h_{A_3}),
\qquad
\end{eqnarray}
\begin{eqnarray}
  \label{eq:dgrs1s1}
  \frac{{d}\Gamma}{{d}w}(\Xi_{QQ_s}\to\Xi_{Q'Q_s})
&=&\frac{G_F^2|V_{QQ'}|^2}{216\pi^3}(w^2-1)^{1/2}(w+1)^3M_F^3
\Biggl\{2(M_I^2+M_F^2-2M_IM_Fw)\cr
&&\times\biggl((2-w)\Bigl[4h_7^2(w)-
(w-1)(h_9(w)+h_{10}(w))^2\Bigr]\cr
&&+\frac{w-1}{w+1}
[h_3(w)+h_4(w)]^2+(w-1)(2h_7(w)+h_9(w)+h_{10}(w))^2
\biggr)\cr
&&+4\Bigl[(M_I-M_F)h_7(w)
-\frac{w-1}{w+1}(M_I+M_F)h_8(w)\Bigr]^2\cr
&&
+\frac{w-1}{w+1}\biggl((w+2)[(M_I+M_F)h_1(w)-(M_I-M_F)h_2(w)]\cr
&&+ (M_I-M_Fw)h_3(w)+(M_F-M_Iw)h_4(w)\cr
&&
+(w^2-1)[M_Fh_5(w)+M_Ih_6(w)]\biggr)^2\Biggr\}
\eta^2(w),
\end{eqnarray}

\begin{eqnarray}
  \label{eq:dgrs1s3}
  \frac{{d}\Gamma}{{d}w}(\Xi_{QQ_s}\to\Xi^*_{Q'Q_s})
&=&\frac{G_F^2|V_{QQ'}|^2}{108\pi^3}(w^2-1)^{1/2}(w+1)^3M_F^3
\Biggl\{(M_I^2+M_F^2-2M_IM_Fw)\cr
&&\times\biggl(h_7^2(w)+[h_7(w)+
(w-1)(h_9(w)+h_{10}(w)]^2\cr
&&+3\left(\frac{w-1}{w+1}\right)^2(h_8^2(w)+[h_8(w)+
(w+1)(h_9(w)-h_{10}(w)]^2)\cr
&&+2\frac{w-1}{w+1}\Bigl(
[h_3(w)-h_4(w)]^2+h_3(w)h_4(w)\cr
&&-(w^2-1)([h_9(w)-h_{10}(w)]^2
+h_9(w)h_{10}(w))\Bigr)
\biggr)\cr
&&+\Bigl[(M_I-M_F)h_7(w)
-\frac{w-1}{w+1}(M_I+M_F)h_8(w)\Bigr]^2\cr
&&
+\frac{w-1}{w+1}\biggl((w-1)[(M_I+M_F)h_1(w)-(M_I-M_F)h_2(w)]\cr
&&+ (M_I-M_Fw)h_3(w)+(M_F-M_Iw)h_4(w)\cr
&&
+(w^2-1)[M_Fh_5(w)+M_Ih_6(w)]\biggr)^2\Biggr\}
\eta^2(w),
\end{eqnarray}
\begin{eqnarray}
  \label{eq:dgrs3s1}
  \frac{{d}\Gamma}{{d}w}(\Xi^*_{QQ_s}\to\Xi_{Q'Q_s})
&=&\frac12 \frac{{d}\Gamma}{{d}w}(\Xi_{QQ_s}\to\Xi^*_{Q'Q_s}),
\end{eqnarray}

\begin{eqnarray}
  \label{eq:dgrs3s3}
  \frac{{d}\Gamma}{{d}w}(\Xi^*_{QQ_s}\to\Xi^*_{Q'Q_s})
&=&\frac{G_F^2|V_{QQ'}|^2}{216\pi^3}(w^2-1)^{1/2}(w+1)^3M_F^3
\Biggl\{(M_I^2+M_F^2-2M_IM_Fw)\cr
&&\times\biggl(5(h_7^2(w)+[h_7(w)+
(w-1)(h_9(w)+h_{10}(w)]^2)\cr
&&+3\left(\frac{w-1}{w+1}\right)^2(h_8^2(w)+[h_8(w)+
(w+1)(h_9(w)-h_{10}(w)]^2)\cr
&&+2\frac{w-1}{w+1}\Bigl(
2[h_3(w)-h_4(w)]^2+5h_3(w)h_4(w)\cr
&&-(w^2-1)(2[h_9(w)-h_{10}(w)]^2
+5h_9(w)h_{10}(w))\Bigr)
\biggr)\cr
&&+5\Bigl[(M_I-M_F)h_7(w)
-\frac{w-1}{w+1}(M_I+M_F)h_8(w)\Bigr]^2\cr
&&
+\frac{w-1}{w+1}\biggl((2w+1)[(M_I+M_F)h_1(w)-(M_I-M_F)h_2(w)]\cr
&&+ (M_I-M_Fw)h_3(w)+(M_F-M_Iw)h_4(w)\cr
&&
+(w^2-1)[M_Fh_5(w)+M_Ih_6(w)]\biggr)^2
+\biggl((M_I-M_Fw)h_3(w)\cr
&&+(M_F-M_Iw)h_4(w)
+(w^2-1)[M_Fh_5(w)+M_Ih_6(w)]\biggr)^2\cr
&&
-2(w+2)(w-1)\biggl((M_I+M_F)h_1(w)\cr
&&-(M_I-M_F)h_2(w)\biggr)^2
\Biggr\}
\eta^2(w).
\end{eqnarray}

\end{document}